\documentclass[preprint,preprintnumbers,amsmath,amssymb]{revtex4}
\usepackage{graphicx}
\usepackage{bm}
\everymath{\displaystyle}
\begin{document}
\title{Potential for measurement of the tensor electric and magnetic
polarizabilities of the deuteron in storage-ring experiments with
polarized beams}

\author{Vladimir G. Baryshevsky and Alexander J. Silenko}

\affiliation{Research Institute for Nuclear Problems, Belarusian
State University, Minsk 220080, Belarus}

\email{bar@inp.minsk.by;  silenko@inp.minsk.by}
\begin{abstract}
Measurement of the tensor electric and magnetic polarizabilities
of the deuteron is of great interest, especially in connection
with the possibilities of COSY and GSI. These polarizabilities can
be measured in storage rings by the frozen spin method providing a
disappearance of $g-2$ precession. This method will be used in the
planned deuteron electric-dipole-moment experiment in storage
rings. The tensor electric polarizability of the deuteron
significantly influences the buildup of the vertical polarization
in the above experiment. The spin interactions depending on the
electric dipole moment, the tensor electric polarizability, and
main systematical errors caused by field misalignments have very
different symmetries. For the considered experimental conditions,
the sensitivity to the deuteron EDM of $1\times10^{-29}~ e\cdot$cm
corresponds to measuring the both of tensor polarizabilities with
an accuracy of
$\delta\alpha_T\approx\delta\beta_T\approx5\times10^{-42}$ cm$^3$.
This conservative estimate can be improved by excluding the
systematical error caused by the field instability which is
negligible for the measurement of the tensor polarizabilities. To
find the tensor magnetic polarizability, the horizontal components
of the polarization vector should be measured.\end{abstract}

\maketitle

\section{Introduction}

Tensor electric and magnetic polarizabilities are important
properties of the deuteron and other nuclei defined by spin
interactions of nucleons. Their measurement provides a good
possibility to examine the theory of spin-dependent nuclear
forces. Methods for determining these important electromagnetic
properties of the deuteron based on the appearance of interactions
quadratic in the spin have been proposed by V. Baryshevsky and
co-workers \cite{Bar1,Bar4,Bar5,Bar3,Barnw,BarJP}. It has been
shown that the tensor polarizabilities cause oscillations of spin
characteristics and transitions between vector and tensor
polarizations of the deuteron. Additional investigations have been
performed in Refs. \cite{PRC,PRC2008,PRC2009}. Interactions
quadratic in the spin and proportional to the tensor electric and
magnetic polarizabilities affect spin dynamics. When an electric
field in the particle rest frame oscillates at the resonant
frequency, an effect similar to the nuclear magnetic resonance
takes place. This effect stimulates the buildup of the vertical
polarization (BVP) of the deuteron beam
\cite{Bar1,Bar4,Bar5,Bar3,Barnw,BarJP}. More general formulas
describing the BVP caused by the tensor electric polarizability of
the deuteron in storage rings have been derived in Ref.
\cite{PRC}. The problem of influence of the tensor electric
polarizability on spin dynamics in such a deuteron
electric-dipole-moment experiment in storage rings has been
investigated \cite{Bar1,Bar4,Bar5,Bar3,Barnw,BarJP,PRC}. It has
been proved that doubling the resonant frequency used in this
experiment dramatically amplifies the effect and provides the
opportunity to make high-precision measurements of the deuteron's
tensor electric polarizability \cite{PRC}.

The tensor magnetic polarizability, $\beta_T$, produces the spin
rotation with two frequencies instead of one, beating with a
frequency proportional to $\beta_T$, and causes transitions
between vector and tensor polarizations
\cite{Bar4,Bar3,Barnw,BarJP}. In Ref. \cite{PRC2008}, the
existence of these effects has been confirmed and a detailed
calculation of deuteron spin dynamics in storage rings has been
carried out. The use of the matrix Hamiltonian derived in Ref.
\cite{PRC} is very helpful for calculating general formulas
describing the evolution of the spin. It is important that the
results obtained by different methods agree.

Significant improvement in the precision of possible experiments
can be achieved if initial deuteron beams are tensor-polarized
\cite{PRC,PRC2008}.

The frozen spin method \cite{NSPC,EDM} provides another
possibility to measure the tensor polarizabilities of the deuteron
and other nuclei. This method ensures that the spin orientation
relative to the momentum direction remains almost unchanged. We
also analyze additional advantages ensured by the use of
tensor-polarized beams and compute the related spin evolution.
Since the spin interactions depending on the EDM, the tensor
electric polarizability, and the main systematical errors have
very different symmetries, all these interactions can be properly
distinguished.

The system of units $\hbar=c=1$ is used.

This is the extended version of the paper prepared for the
Proceedings of 19th International Spin Physics Symposium
(September 27 -- October 2, 2010, Julich, Germany).

\section{General equations}

We use the matrix Hamiltonian equation and the matrix Hamiltonian
$H$ for determining the evolution of the spin wave function:
\begin{equation}
 \begin{array}{c}
 i \frac{d\Psi}{dt}=H\Psi, ~~~ \Psi=\left(\begin{array}{c}
C_{1}(t)
 \\ C_{0}(t) \\ C_{-1}(t) \end{array}\right).
 \end{array}\label{eq19t}\end{equation}

The three-component wave function $\Psi$, which is similar to a
spinor, consists of the amplitudes $C_{i}(t)$ characterizing
states with definite spin projections onto the preferential
direction ($z$ axis). Correction to the Hamilton operator caused
by the tensor polarizabilities has the form \cite{PRC}
\begin{equation} \begin{array}{c}
V=-\frac{\alpha_T}{\gamma}(\bm S\cdot\bm
E')^2-\frac{\beta_T}{\gamma}(\bm S\cdot\bm B')^2,
\end{array} \label{eqTre} \end{equation}
where $\alpha_T$ is the tensor electric polarizability, $\gamma$
is the Lorentz factor, and $\bm E'$ and $\bm B'$ are the electric
and magnetic fields in the rest frame of the deuteron. Let us note
that the components of $\bm E'$ and $\bm B'$ orthogonal to the
deuteron momentum contain the Lorentz factor (see below). As a
result, the effect does not decrease at high energies
\cite{Bar1,Bar4,Bar5,Bar3,Barnw,BarJP}.

The spin motion in storage rings is measured relative to the axes
of the cylindrical coordinate system. Therefore, cylindrical
coordinates are used in the present work. The horizontal axes of
the cylindrical coordinate system are connected with the position
of the particle and rotate at the instantaneous frequency of its
revolution.
The description of spin effects in the cylindrical coordinate
system strongly correlates with that in the frame rotating at the
instantaneous frequency of orbital revolution of the deuteron. The
frequency of spin precession in the rotating frame coincides with
that in the cylindrical coordinate system (see Refs.
\cite{PRC,PRC2008,PRC2009}).

In the rotating frame, the motion of deuterons is relatively slow
because it can be caused only by beam oscillations and other
deflections from the ideal trajectory.

The Hamiltonian operator is defined by \cite{PRC}
\begin{equation} \begin{array}{c}
{\cal H}={\cal H}_0+\bm S\cdot\bm\omega_a+V,
\end{array} \label{eqapp} \end{equation}
where $\bm\omega_a$ is the angular velocity of the spin precession
relatively to the momentum direction ($g-2$ precession).

In the considered case, the expressions for $\bm E'$ and $\bm B'$
in terms of the unprimed laboratory fields have the form
\begin{equation} \begin{array}{c}
\bm E'=\gamma(E_\rho+\beta_\phi B_z)\bm e_\rho, ~~~ \bm
B'=\gamma(\beta_\phi E_\rho+B_z)\bm e_z,
\end{array} \label{eqTr} \end{equation}
where $\beta_\phi=\bm\beta\cdot \bm e_\phi\equiv \bm v\cdot \bm
e_\phi/c$.

When the frozen spin method is used, the quantity $\bm\omega_a$ is
very small and the fields satisfy the following relation
\cite{EDM}:
\begin{equation} \begin{array}{c}
E_\rho=\frac{a\beta_\phi\gamma^2}{1-a\beta^2\gamma^2} B_z.
\end{array} \label{eqc} \end{equation}
Since the main electric field is radial and almost orthogonal to
the particle (nucleus) trajectory, its effect on a change of the
$\gamma$ factor can be neglected. This factor can be supposed to
be constant.

Therefore,
\begin{equation} \begin{array}{c}
V=-\frac{\gamma
B_z^2}{(1-a\beta^2\gamma^2)^2}\left[\alpha_T(1+a)^2\beta^2
S_\rho^2+\beta_TS_z^2\right].
\end{array} \label{eqTrI} \end{equation}

The matrix Hamiltonian has the form \cite{PRC,PRC2009}
\begin{equation}
 \begin{array}{c}
 H=\left(\begin{array}{ccc} E_{0}\!+\!\omega_0\!+\!{\cal A}\!+\!{\cal B} & 0 & {\cal A} \\
0 & E_{0}\!+\!2{\cal A} & 0 \\
{\cal A} & 0 & E_{0}\!-\!\omega_0\!+\!{\cal A}\!+\!{\cal B}
 \end{array}\right),
 \end{array}\label{eqMH}\end{equation}
where $E_0$ is the zero energy level,
$\omega_0=\left(\omega_a\right)_z$,
\begin{equation}
\begin{array}{c}
{\cal A}=-\alpha_T\frac{(1+a)^2\beta^2\gamma
B_z^2}{2(1-a\beta^2\gamma^2)^2}, ~~~ {\cal B}=-\beta_T\frac{\gamma
B_z^2}{(1-a\beta^2\gamma^2)^2}.
\end{array}\label{eqMHt}\end{equation}

Equations (\ref{eqMH}) and (\ref{eqMHt}) are basic equations
defining the dynamics of the deuteron spin in storage rings when
the frozen spin method is used \cite{PRC2009}.

We are interested in the case when the particle or nucleus has a
fixed spin projection ($S_l=+1,0$, or $-1$) onto the certain
direction $\bm l$ defined by the spherical angles $\theta$ and
$\psi$. The azimuth $\psi$ is determined in relation to the
cylindrical axes $\bm e_\rho$ and $\bm e_\phi$. The $\psi=0$ case
characterizes the spin directed radially outward.

The polarization of particles (nuclei) is described by the
three-component polarization vector $\bm P$ and the polarization
tensor $P_{ij}$, which has five independent components.

\section{Evolution of vector polarization of the deuteron beam}

In Ref. \cite{PRC2008}, off-diagonal components of the Hamiltonian
(\ref{eqMHt}) were not taken into account, because their effect on
the rotating spin did not satisfy the resonance condition. These
components cannot, however, be neglected in the considered case
because the resonant frequency $\omega_0$ can be very small.

The best conditions for a measurement of the tensor
polarizabilities of the deuteron and other nuclei can be achieved
with the use of tensor-polarized initial beams. In this case, we
may confine ourselves to the consideration of a zero projection of
the deuteron spin onto the preferential direction. When this
direction is defined by the spherical angles $\theta$ and $\psi$,
the general equation describing the evolution of the polarization
vector has the form  \cite{PRC2009}
\begin{equation}
\begin{array}{c}
P_\rho(t)=\sin{(2\theta)}\biggl\{\Bigl[\cos{(\omega't)}\sin{\psi}
+\frac{\omega_0}{\omega'}\sin{(\omega't)}\cos{\psi}\Bigr]\sin{(bt)}\!+\!
\frac{{\cal A}}{\omega'}\sin{(\omega't)}\cos{(bt)}\sin{\psi}\biggr\},\\
P_\phi(t)=\sin{(2\theta)}\biggl\{\Bigl[-\cos{(\omega't)}\cos{\psi}
+\frac{\omega_0}{\omega'}\sin{(\omega't)}\sin{\psi}\Bigr]\sin{(bt)}
\!+\!\frac{{\cal A}}{\omega'}\sin{(\omega't)}\cos{(bt)}\cos{\psi}\biggr\},\\
P_{z}(t)=-\frac{2{\cal
A}}{\omega'}\sin^2{\theta}\sin{(\omega't)}\Bigl[\cos{(\omega't)}
\sin{(2\psi)} 
+\frac{\omega_0}{\omega'}\sin{(\omega't)}\cos{(2\psi)}\Bigr],
\end{array}
\label{prop}
\end{equation}
where
\begin{equation}
\omega'=\sqrt{\omega_0^2+{\cal A}^2}, ~~~ b={\cal B}-{\cal A}.
\label{eqb}
\end{equation}
When the frozen spin method is used,
\begin{equation}
b=-\frac{\gamma
B_z^2}{(1-a\beta^2\gamma^2)^2}\left[\beta_T-\frac{1}{2}\alpha_T(1+a)^2\beta^2\right].
\label{eqbf}
\end{equation}

As a rule, we can neglect ${\cal A}^2$ as compared with
$\omega_0^2$ and use the approximation $|b|t\ll1$. In this case
\cite{PRC2009}
\begin{equation}
\begin{array}{c}
P_\rho(t)=\sin{(2\theta)}\left[bt\sin{(\omega_0t+\psi)}+ 
\frac{{\cal A}}{\omega_0}\sin{(\omega_0t)}\sin{\psi}\right],\\
P_\phi(t)=\sin{(2\theta)}\!\left[-bt\cos{(\omega_0t+\psi)}\!+\!
\frac{{\cal A}}{\omega_0}\sin{(\omega_0t)}\cos{\psi}\right],\\
P_{z}(t)=-\frac{2{\cal
A}}{\omega_0}\sin^2{\theta}\sin{(\omega_0t)}\sin{(\omega_0t+2\psi)}.
\end{array}
\label{prp}
\end{equation}

When the initial deuteron beam is vector-polarized, the direction
of its polarization can be also defined by the spherical angles
$\theta$ and $\psi$. Such a polarization (with $\theta=\pi/2$)
will be used in the planned deuteron electric-dipole-moment (EDM)
experiment \cite{dEDM}. The EDM manifests in
an appearance of a 
vertical component of the polarization vector.

The evolution of this component defined by the tensor
polarizabilities of the deuteron is given by
\begin{equation}
\begin{array}{c}
P_{z}(t)=\left[1-\frac{2{\cal
A}^2}{{\omega'}^2}\sin^2{(\omega't)}\right]\cos{\theta}
+\frac{{\cal
A}}{\omega'}\sin^2{\theta}\sin{(\omega't)}\Bigl[\cos{(\omega't)}
\sin{(2\psi)} \\
+\frac{\omega_0}{\omega'}\sin{(\omega't)}\cos{(2\psi)}\Bigr].
\end{array}
\label{propv}
\end{equation}
The tensor magnetic polarizability does not influence on $P_{z}$.

In the same approximation as before,
\begin{equation}
\begin{array}{c}
P_{z}(t)=\cos\theta+\frac{{\cal
A}}{\omega_0}\sin^2{\theta}\sin{(\omega_0t)}\sin{(\omega_0t+2\psi)}.
\end{array}
\label{aprop}
\end{equation}

\section{Distinguishing features of the tensor
polarizabilities and electric-dipole-moment effects}

The tensor electric polarizability can in principle imitate the
presence of the EDM because both of them stimulate the buildup of
the vertical polarization of the deuteron beam. The exact equation
of spin motion
with allowance for the EDM 
has been obtained in Ref. \cite{NSPC} specifically for the EDM
experiment. In the considered case, the angular velocity of spin
rotation is equal to
\begin{eqnarray}
\bm\omega_a=\omega_0\bm e_z+{\cal C}\bm e_\rho, ~~~ {\cal
C}=-\frac{e\eta}{2m}\cdot\frac{1+a}{1-a\beta^2\gamma^2}\beta_\phi
B_z, \label{eq8}\end{eqnarray} where $\eta=2dm/(eS)$ is the factor
similar to the $g$ factor for the magnetic moment with $d$ being
the EDM.

When the tensor polarizabilities are not taken into account, the
spin rotates about the direction
$$\bm e_z'=\frac{{\cal C}}{\omega'}\bm e_\rho+\frac{\omega_0}{\omega'}\bm e_z$$
with the angular frequency $\omega'=\sqrt{\omega_0^2+{\cal C}^2}$.

When the initial beam is vector-polarized, the final polarization
vector is defined by \cite{PRC2009}
\begin{equation}
\begin{array}{c}
P_{\rho}(t)=\frac{{\omega_0\cal
C}}{{\omega'}^2}\left[1-\cos{(\omega't)}\right]\cos{\theta} 
+\!\left[1\!-\!\frac{2\omega_0^2}{{\omega'}^2}\sin^2{\frac{\omega't}{2}}\right]\sin{\theta}
\cos{\psi}-\frac{\omega_0}{\omega'}\sin{(\omega't)}\sin{\theta}\sin{\psi},\\
P_{\phi}(t)=\sin{(\omega't)}\left(\frac{\omega_0}{\omega'}
\sin{\theta}\cos{\psi}-\frac{{\cal
C}}{\omega'}\cos{\theta}\right) 
+\cos{(\omega't)}
\sin{\theta}\sin{\psi},\\
P_{z}(t)=\left[1-\frac{2{\cal
C}^2}{{\omega'}^2}\sin^2{\frac{\omega't}{2}}\right]\cos{\theta}
+\frac{{\omega_0\cal
C}}{{\omega'}^2}\left[1-\cos{(\omega't)}\right]\sin{\theta}
\cos{\psi}+\frac{{\cal
C}}{\omega'}\sin{(\omega't)}\sin{\theta}\sin{\psi}.
\end{array}
\label{propedm}
\end{equation}
If we neglect terms of the order of ${\cal C}^2$, the vertical
component of the polarization vector takes the form
\begin{equation}
\begin{array}{c}
P_{z}(t)=\cos{\theta}+\frac{2{\cal
C}}{\omega_0}\sin{\theta}\sin{\frac{\omega_0t}{2}}\sin{\frac{\omega_0t+2\psi}{2}}.
\end{array}
\label{appredm}
\end{equation}

Although Eqs. (\ref{aprop}) and (\ref{appredm}) are similar, the
effects of the tensor electric polarizability and the EDM have
different angular dependencies and can be properly separated.

Carrying out the dEDM experiment in storage rings \cite{dEDM}
ensures the wonderful possibility of measurement of the tensor
electric and magnetic polarizabilities. The EDM collaboration
plans to use the frozen spin method and to have multiple deuteron
beam bunches circulating in the storage ring at the same time with
opposite states of polarization. Main systematical errors caused
by field misalignments will be canceled by clockwise (CW) and
counterclockwise (CCW) consecutive beam injections \cite{dEDM}.
Initial beam polarization will be horizontal and the final
vertical polarization will be measured. Spin interactions
depending on the EDM, the tensor electric polarizability, and the
main systematical errors have very different symmetries. In the
Table \ref{table}, we show the behavior of the final vertical
polarization caused by these spin interactions when the direction
of either the initial polarization or the beam is reversed. The
plus and minus signs denote reversing and conserving the final
vertical polarization, respectively. In the dEDM experiment, one
will obtain four blocks of data for two opposite states of initial
polarization and two beam directions. Evidently, the tensor
electric polarizability can be determined with summing up the data
for two opposite states of initial polarization. To find the
tensor magnetic polarizability of the deuteron, the horizontal
components of the polarization vector should be measured.
Determination of the deuteron EDM can be made with calculating the
difference of the results for two opposite beam directions and
taking into account the correction for the tensor electric
polarizability.

\begin{table}[htp]\centering
\caption{Reversing ($+$) or conserving ($-$) the final vertical
polarization when either the initial polarization or the direction
of the beam is reversed.}
\begin{tabular}{|c|c|c|c}
\hline
    Spin interactions  &  Reversing the       & Reversing the \\
                       & initial polarization & beam direction \\
\hline
EDM &     $+$    &        $-$    \\
\hline
Tensor electric polarizability &     $-$    &     $-$      \\
\hline
Systematical errors caused &    $+$     &    $+$      \\
by field misalignments &         &          \\
\hline
\end{tabular}
\label{table}\end{table}

\section{Discussion and summary}

Experimental conditions needed for the measurement of the tensor
polarizabilities and the EDMs of nuclei in storage rings
\cite{EDM,dEDM} are similar. Equation (\ref{eqc}) shows that the
radial electric field should be sufficiently strong to eliminate
the effect of the vertical magnetic field on the spin. As a
result, the frozen spin method provides a weaker magnetic field
than other methods. This factor is negative because the evolution
of the spin caused by both the tensor polarizabilities and the
EDMs strongly depends on $B_z$. Nevertheless, the Storage Ring EDM
Collaboration considers the frozen spin method to be capable of
detecting the deuteron EDM of the order of $10^{-29}~e \cdot\,$cm.
Another method for searching for the deuteron EDM in storage rings
is the resonance method developed in Ref. \cite{OMS}. This method
is based on a strong vertical magnetic field and an oscillatory
resonant longitudinal electric field. Both methods can be
successfully used for the measurement of the tensor electric and
magnetic polarizabilities of the deuteron and other nuclei
\cite{Bar1,Bar4,Bar5,Bar3,Barnw,BarJP,PRC}.

We can evaluate the precision of measurement of the tensor
polarizabilities of the deuteron via its comparison with the
expected sensitivity of the deuteron EDM experiment. For the
considered experimental conditions \cite{dEDM}, the sensitivity to
the EDM of $1\times10^{-29} ~e \cdot\,$cm corresponds to measuring
the tensor electric polarizability with an accuracy of
$\delta\alpha_T\approx 5\times10^{-42}$ cm$^3$.

There are three independent theoretical predictions for the value
of the tensor electric polarizability of the deuteron, namely
$\alpha_T=-6.2\times10^{-41}$ cm$^3$ \cite{CGS},
$-6.8\times10^{-41}$ cm$^3$ \cite{JL}, and $3.2\times10^{-41}$
cm$^3$ \cite{FP}. The first two values are very close to each
other but they do not agree with the last result. The theoretical
estimate for the tensor magnetic polarizability of deuteron is
$\beta_T=1.95\times10^{-40}$ cm$^3$ \cite{CGS,JL}.

We can therefore conclude that the expected sensitivity of the
deuteron EDM experiment allows us to measure the tensor electric
polarizability with an absolute precision of
$\delta\alpha_T\approx5\times10^{-42}$ cm$^3$ which corresponds to
the relative precision of the order of $10^{-1}$. This estimate
made for the vector-polarized initial beam is rather conservative
and can be improved by excluding the systematical error caused by
the field instability which is negligible for the measurement of
the tensor polarizabilities. The best sensitivity in the
measurement of $\alpha_T$ can be achieved with the use of a
tensor-polarized initial beam. When the vector polarization of
such a beam is zero, spin rotation does not occur. In this case,
there are no related systematic errors caused by the radial
magnetic field or other factors. In the general case, such
systematic errors are proportional to a residual vector
polarization of the beam. This advantage leads to a sufficient
increase in experimental accuracy \cite{PRC,PRC2008}. In this
case, our preliminary estimate of experimental accuracy is
$\delta\alpha_T\sim 10^{-43}$ cm$^3$.

The frozen spin method can also be successively used for the
measurement of the tensor magnetic polarizability. Equations
(\ref{prop})--(\ref{prp}) show that the preferential direction of
initial tensor polarization is defined by $\theta=\pi/2$ and
$\theta=\pi/4$ for measuring the tensor electric and magnetic
polarizabilities, respectively. In the latter case, the horizontal
components of the polarization vector should be measured. Owing to
a restriction of spin rotation in the horizontal plane, the
achievable absolute precision of measurement of the tensor
magnetic polarizability of the deuteron is of the same order
($\delta\beta_T\sim 10^{-43}$ cm$^3$). A comparison with the
theoretical estimate \cite{CGS,JL} shows that the relative
precision of measurement of this quantity can be rather high
($\delta\beta_T/\beta_T\sim 10^{-3}$).

All the formulas derived here are applicable to any spin-1
nucleus. Moreover, the evolution of the polarization vector
defined by spin tensor effects has to be similar for nuclei with
any spin $S\geq1$ despite difference of spin matrices and has to
agree with classical spin physics. The description of dynamics of
the vector and tensor polarizations for arbitrary $S$ and the
investigation of special features of higher spins have been
carried out in Ref. \cite{matd}.

Thus, the frozen spin method can be effectively used for the
high-precision determination of the tensor electric and magnetic
polarizabilities of the deuteron and other nuclei. The experiments
could be carried out at COSY and GSI.

\section*{Acknowledgments}

This work was supported by the Belarusian Republican Foundation
for Fundamental Research (Grant No. $\Phi$10D-001).


\end{document}